\documentclass[twocolumn,prb,showpacs,preprintnumbers,amsmath,amssymb]{revtex4}

\usepackage[dvips]{graphicx}
\usepackage{mathptmx}
\usepackage{verbatim}

\begin{document}
\title{Lattice Theory of Pseudospin Ferromagnetism in Bilayer Graphene:\\
 Competing Orders and Interaction Induced Quantum Hall States}
\author{Jeil Jung} \email{jeil@physics.utexas.edu}
\affiliation{Department of Physics, University of Texas at Austin, USA}
\author{Fan Zhang} 
\affiliation{Department of Physics, University of Texas at Austin, USA}
\author{Allan H. MacDonald}
\affiliation{Department of Physics, University of Texas at Austin, USA}
\date{\today{}}
\begin{abstract}
In mean-field-theory bilayer graphene's massive Dirac fermion model has a family of broken inversion symmetry ground states
with charge gaps and flavor dependent spontaneous inter layer charge transfers.  
We use a lattice Hartree-Fock model to explore some of the physics which controls whether or not this type of broken 
symmetry state, which can be viewed as a pseudospin ferromagnet,  
occurs in nature.  
We find that inversion symmetry is still broken in the lattice model 
and estimate that transferred areal densities are $\sim 10^{-5}$ electrons per carbon atom, 
that the associated energy gaps are $\sim 10^{-2} {\rm eV}$,  that the 
ordering condensation energies are $\sim  10^{-7} eV$ per carbon atom,
and that the energy differences between competing orders at the neutrality point are 
 $\sim 10^{-9} eV$ per carbon atom.  
We explore the quantum phase transitions induced 
by external magnetic fields and by externally controlled electric potential differences
between the layers.  
We find, in particular, that in an external magnetic field
coupling to spontaneous orbital moments favors broken time-reversal-symmetry states that 
have spontaneous quantized anomalous Hall effects.   
Our theory predicts a non monotonic behavior of the band gap at neutrality as a function 
of interlayer potential difference 
in qualitative agreement with recent experiments.
\end{abstract}

\maketitle

\section{Introduction}
Recent experimental progress 
\cite{geim,kim}
in isolating and measuring the electronic properties of graphene single and multilayers 
has opened a new topic in two-dimensional electron system (2DES) physics \cite{geim_macdonald}. 
Electronic wavefunctions in graphene systems are often described using a pseudospin language
in which the spinors specify wavefunction components on different sublattices.
Although the properties of graphene 2DES's can often be successfully described using an effective  
non-interacting electron model, studies of 
electron-electron interactions effects have revealed some qualitative 
differences compared to ordinary 2DES's\cite{yafis,chiral_2d} that are related to these sublattice pseudospin 
degrees-of-freedom.  

In the case of AB stacked bilayer graphene, there are four C sites and four $\pi$-orbitals per
unit cell, but two of these are repelled from the neutral system Fermi level by interlayer hopping. 
This circumstance leads to a low-energy massive chiral fermion model\cite{falko} with two-component spinors, and a 
crystal-momentum $\vec{p}$ dependent pseudo-magnetic field with a magnitude that 
varies as $p^2$ and an orientation angle twice as large as the momentum orientation 
angle $\phi_{\vec{p}}$.  Neutral system states have one occupied pseudospin for
each distinct set of momentum, spin, and valley labels.  Recently\cite{pmag} Min {\em et al.} pointed out that when 
Coulombic electron-electron interactions are added to the massive chiral fermion model, 
the mean-field theory ground state pseudospins of each spin-valley flavor break symmetry by rotating out 
of the $x-y$ plane, developing $\hat{z}$ components with a common spontaneously chosen sign
and magnitudes which are larger at small $p$.
Because the two-sublattices from which the two-band-model pseudospins are constructed are located in 
opposite layers, the broken symmetry transfers charge between layers. 
It is therefore characterized in momentum space by a vortex with 
vorticity $v=2$ and a flavor-dependent core polarized along one of the polar directions, and in real space by
flavor-dependent  
uniform layer polarization.  Different members of the family of states are distinguished by the spin-valley flavor
dependence of the sense of layer pseudospin orientation in the momentum-space vortex cores. 
The origin of the broken symmetry,
which we refer to here as pseudospin ferromagnetism, \cite{pmag} 
is the $p^2$ pseudospin-splitting at small $p$, which leads to infrared divergences \cite{fanzhang}
in particle-hole polarization loops,
combined with the frustrating effect of pseudospin chirality which leads to relatively
stronger exchange interactions for $\hat{z}$-polarized pseudospins.  

The broken symmetry states can be classified\cite{nandkishore2, xiaofan} in a manner which highlights their valley
and spin dependent momentum space Berry curvatures, 
which are in turn closely related\cite{niureview,NagaosaReview} to the 
Hall responses of the system.
Indeed, recent experiments\cite{yacoby_latest} demonstrate that bilayer graphene  
exhibits a quantized quantum Hall effect in the absence of an external magnetic field.
Over and above the basic science interest deriving from this presently unique property, 
it is possible that pseudospin order in bilayer graphene could play a role in graphene electronics\cite{bilayerpseudospintronics} by 
introducing hysteresis and dramatically enhancing the influence of gates on conductance.  

In this article we report on a study of pseudospin ferromagnetism
in a $\pi$-orbital tight-binding model for bilayer graphene, which is conveniently able to 
capture and establish the role of some bilayer graphene $\pi$-band features neglected in the massive chiral fermion model. 
We estimate that the density shift for each flavor is $\sim 10^{-5}$ electrons per carbon atom,
that the gaps are $\sim 10^{-2} eV$, that the total condensation energy  
is $ \sim 10^{-7} \,\, eV$ per carbon atom, and that the energy differences between competing
ordered states is $\sim 10^{-9} \,\,eV$ per carbon atom. 
The energy differences between competing ordered states,
which are between one to two orders or magnitude 
smaller than the condensation energy, are sensitive to 
roughly estimated lattice scale details
of the model we employ.
The competing states are classified in the first place as either anomalous Hall states, in which opposite valleys are polarized toward opposite layers, or valley Hall states in which they are polarized toward the same layer.  
We find that valley Hall states have slightly lower energies.  Because anomalous Hall states have spontaneous orbital magnetism they are favored by external magnetic fields.  We estimate that, because the energy differences between states are very small, extremely weak fields are sufficient to induced phase transitions between valley Hall and anomalous Hall states.  On the other hand potential differences between layers favor valley Hall states.  When the spin degree of freedom is accounted for we find that the spontaneous energy gap in neutral bilayers first decreases and then increases with potential difference, in qualitative agreement with experiment.

In the following two sections we explain the model Hamiltonian we employ and some of the technical 
details of the calculations we have carried out. 
The main results are presented in the Section IV where we introduce the topological 
classification of solutions and use it to discuss the properties of broken symmetry solutions
for vanishing, moderate, and strong externally controlled electric potential differences between the layers.
(Below we refer to interlayer potential differences, which always play a key role in 
graphene bilayer physics, as potential biases.)
We also discuss coupling to an external magnetic field which can drive transitions 
between competing states, some of which have broken time-reversal-symmetry with 
associated anomalous Hall effects and orbital magnetism.

We close the paper with a summary and some suggestions for further research.

\section{Four-Band $\pi$-orbital Tight-Binding Model of Bilayer Graphene}
We describe bilayer graphene using a lattice model with one atomic $2 p_z$ orbital per carbon site.
We write the model's Bloch basis states in the form
\begin{equation}
\label{basis}
\psi_{{\bf k}\kappa} \left({\mathbf r} \right) = \frac{1}{\sqrt{N}} \sum_{i} e^{i {\mathbf k} \left( {\mathbf R}_i + {\bf \tau}_{\kappa} \right)} 
\phi \left({\mathbf r} - {\mathbf R}_i - {\bf \tau}_{\kappa} \right),
\end{equation}
where $N$ is the total number of unit cells in the system, $\phi  \left({\mathbf r} \right)$ is the band's 
Wannier wavefunction, and $\kappa$ labels the carbon site with position 
${\bf \tau}_{\kappa}$ relative to a the triangular lattice vector ${\mathbf R}_i$.
(We comment later on the possible role of screening effects from
the $p$ and $s$ orbitals which form the $\sigma$ and $\sigma^*$ bonds
neglected in this model.)
Following the convention used in Refs.[\onlinecite{falko,hongki}],
we use the notations {\em A}, {\em B}, $\tilde{A}$, $\tilde{B}$ for the four sublattice indexes $\kappa$,
where {\em B} and $\tilde{A}$ are the opposite-layer near-neighbor-pair sites. 
With this convention, the four band tight-binding model Hamiltonian of a graphene bilayer is:
\begin{equation}
\label{hamil}
{H}_0= 
\begin{pmatrix}
     0 & \gamma_0 f & \gamma_4 f & \gamma_3 f^*   \\
\gamma_0 f^*    & 0 & \gamma_1 & \gamma_4 f  \\
\gamma_4 f^*   & \gamma_1 & 0 &  \gamma_0 f  \\
 \gamma_3 f & \gamma_4 f^* & \gamma_0 f^* & 0
\end{pmatrix}
\end{equation}
where
\begin{eqnarray}
f\left( {\bf k} \right) &=&   e^{ i k_y a / \sqrt{3} } \left( 1 + 2 e^{-i 3 k_{y} a / 2\sqrt{3}}  
                                       \cos \left(  \frac{k_x a}{2} \right)    \right)  
\end{eqnarray}
with $a = 2.46 \AA$ arises from a sum over the three near-neighbor hops within a layer. 
We have neglected differences in on-site energies and next nearest neighbor hopping processes 
which give rise to electron-hole asymmetry and do not play an important role in pseudospin ferromagnetism.
The tight-binding model parameters $\gamma_i$ should not be confused with the 
Slonczewski-Weiss, McClure \cite{swm} model parameters for bulk graphite,
despite the obvious similarities in notation. 
In our calculations we adopt conventions similar to those of Ref.[\onlinecite{gamma3}]
for bilayer graphene, taking the values $\gamma_0 = -3.12 \,\, eV$, $\gamma_1 = -0.377$,
$\gamma_3 = -0.29 \,\, eV$ and $\gamma_4 = -0.12 \,\, eV$ for the hopping parameters.
Only the intralayer nearest neighbor ($\gamma_0$) process and interlayer tunneling ($\gamma_1$) 
process are retained in the minimal tight-binding model. 
The trigonal warping ($\gamma_3$) process which connects the $A$ and $\widetilde{B}$ sites 
is responsible for the leading circular symmetry breaking near the 
valley points,
while the ($\gamma_4$) process which connects $A$ and $\widetilde{A}$ sites 
influences the intralayer charge imbalance between sublattices $A$ and $B$.

\section{Lattice Model Mean-Field Theory} 
Because of the importance\cite{chiral_2d} of non-local exchange in graphene systems,
a Hartree-Fock mean-field theory approximation \cite{ostlund} is a natural first step in considering electron-electron interaction effects.
When Coulomb interactions are added to the $\pi$-band tight-binding model the interaction terms in the mean-field
Hamiltonian take the form:
\begin{eqnarray}
\label{hartreefock}
V^{HF} = \sum_{{\bf k} \lambda \lambda'} U_H^{\lambda \lambda'} N_{\lambda'} c_{{\bf k} \lambda}^{\dag} c_{{\bf k} \lambda} 
-W^{X}_{{\bf k} {\lambda' \lambda}}  c_{{\bf k} \lambda}^{\dag} c_{{\bf k} \lambda'} 
\end{eqnarray}
where $\lambda$ is a composite label for sublattice $\kappa$ and spin $\sigma$.  The first term on the right hand side
of Eq.~(\ref{hartreefock}) is the Hartree term:
\begin{eqnarray}
N_{\lambda} &=& N_{\kappa \sigma} =\sum_{{\bf k}'} \left< c^{\dag}_{{\bf k}' \lambda} c_{{\bf k}' \lambda} \right>  = \sum_{{\bf k}'} n_{{\bf k}' \lambda} \\
U_H^{\lambda \lambda'} &=&  \frac{\delta_{\sigma,\sigma'}}{A} \sum_{\bf G} \exp{\left[ i {\bf G} \left(\tau_{\kappa} - \tau_{\kappa^{\prime}} \right) \right]} 
\left| \widetilde{f}\left( \left| {\bf G}   \right|\right) \right|^2 V^{ \kappa \kappa^{\prime}} \left( \left|  {\bf G}   \right|  \right), \,\,\,\, \quad,
\label{uhkern}
\end{eqnarray}
where ${\bf G}$ is a reciprocal lattice vector.  The second is the Fock (exchange) term: 
\begin{eqnarray}
W^{X}_{{\bf k} \lambda \lambda'} &=&
 \sum_{{\bf k}'} U_X^{\kappa \kappa'} 
 \left( {\bf k}' - {\bf k} \right) 
 \left<c_{{\bf k}' \lambda'}^{\dag} c_{{\bf k}' \lambda} \right> \\ 
U_X^{\kappa \, \kappa^{\prime}} \left( {\bf q} \right) &=& \frac{1}{A} \sum_{\bf G} \; \exp{ \left[ i {\bf G} \left(\tau_{\kappa} - \tau_{\kappa^{\prime}} \right) \right]} 
\nonumber \\
&\times& \left| \widetilde{f}\left( \left| {\bf q}  -  {\bf G}  \right|\right) \right|^2 V^{ \kappa \kappa^{\prime}} \left( \left|  {\bf q} - {\bf G}  \right|  \right).
\label{uxkern}
\end{eqnarray}
In Eq.(\ref{uhkern}) and Eq.(\ref{uxkern}) the two-dimensional Coulomb interaction
$V^{\kappa \kappa^{\prime}} \left( {\bf q} \right) =  2 \pi e^2 / \left(  \left| \bf q \right|   \epsilon_r \right)  $
when $\kappa$ and $\kappa^{\prime}$ refer to the same layer and 
$\left( 2 \pi e^2 /\left( { \left| \bf q \right|}   \epsilon_r \right)  \right)  \exp{ \left[  - \left| {\bf q} \right|  c \right] } $
when $\kappa$ and $\kappa^{\prime}$ refer to the opposite layers.
Here $\epsilon_r$ is the relative dielectric constant, 
$c=3.35 \AA$ is the interlayer separation,
$A$ is the total area of the graphene sheet, and we use  
\begin{equation}
\widetilde{f}\left( q \right) = (1 -  \left(r_0 q\right)^2 ) / (  (1 + \left(r_0 q \right)^2 )^4  )
\label{formfactor}
\end{equation}
as a form factor which accounts for the spread of the $\pi$-orbital charge on each site.
This simple form assumes an isotropic site-localized charge distribution.
Eq.~(\ref{formfactor}) was obtained by Fourier transforming the radial charge distribution of a hydrogenic $2 p$ orbital. 
The use of $r_0 = \widetilde{a}_0 = a_0/ \sqrt{30}$ would yield a root mean square radius
corresponding to the covalent radius of the carbon atom $a_0 = 0.77 \AA$. 
If we consider screening from the $\sigma$ band electrons neglected in our model and 
the fact that the charge density distribution of a $p_z$ orbital is far from spherical
we expect that larger values of $r_0$, which effectively 
reduce onsite repulsion, would be more appropriate.
For most of our calculations we have therefore used the value $r_0 = 3 \widetilde{a}_0$.

As explained in the introduction, pseudospin ferromagnetism in bilayer graphene 
can be neatly described using the two-band massive chiral fermion model. 
This approach has two shortcomings which the present calculation is intended to 
alleviate.  First of all, the model has to rely on a crude ultraviolet-cutoff to
account for the limited range of energy $\sim \gamma_1$ over which 
it is applicable.  At moderate interaction strengths the amount of 
charge transferred between layers determined in the massive chiral 
fermion model calculation is strongly influenced by this cut-off. 
Secondly, the calculation described in Ref. (\onlinecite{pmag}) 
relies on the model's circular symmetry for a number of simplifications.
When direct hopping between the low-energy $A$ and $\tilde{B}$ sites, the 
$\gamma_3$ process, is included in the Hamiltonian the model's Fermi lines are 
no-longer circular and the continuum model loses some of its attractive simplicity.
\cite{trigonal1,trigonal2}
These processes are known to be essential at 
very weak interaction strengths since they remove the 
infrared divergences\cite{fanzhang,yang} responsible for the instabilities of the massive 
chiral fermion model.  
The present lattice model reduces to the continuum model
at low energies, accounts naturally for the limited validity range of two-band models by
retaining all four $\pi$-bands, and can deal with the loss of circular symmetry without 
any additional complication.  

The main challenges which arises in 
practical implementation of the lattice model mean-field-theory 
lie in the numerical optimization of a problem in which the small portion of
the Brillouin-zone close to one of the valley points plays a dominating role 
and must be sampled densely. 
\begin{figure}[htbp]
\begin{center}
\includegraphics[width=6.6cm,angle=90]{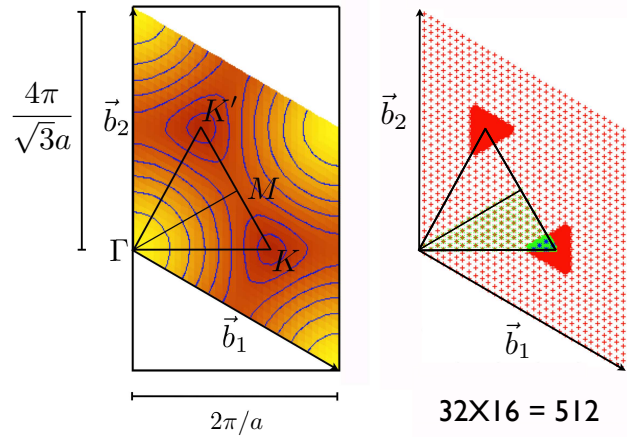} 
\caption{
{\em Left panel:}
Band structure of graphene represented in the 
primitive zone defined by the reciprocal lattice vectors $\vec{b}_1$ and $\vec{b}_2$
of the honeycomb lattice.
The equilateral triangle represents the region in the primitive cell we need to sample
in order to fully describe a system in which the symmetry between $K$ and $K'$ valleys
is broken.  The smaller triangles enclose inequivalent regions in $k$-space
associated with $K$ and $K'$ valleys.
{\em Right panel:}
An example of coarse sampling of $k$-points in the primitive cell 
supplemented by a finer grid
in hexagons generated around those coarse $k$-points
near the $K$ and $K'$ valleys. The illustrated example consists of $32 \times 32$
coarse points and a $16$-fold enhancement of density resulting in an effective
sampling of $512 \times 512$ points in the primitive cell near the valleys.
}
\label{primitive}
\end{center}
\end{figure}
It is essential that we have sufficiently dense ${ k}$-point sampling
near the Dirac points, and at the same time sample the full Brillouin zone.
The increase of computational load with ${ k}$-point sampling density is 
particularly rapid in the Hartree-Fock calculations because the Hartree-Fock 
matrix element at one {$ k$}-point depends on the occupied wavefunctions at all other {$ k$}-points. 
In Fig. \ref{primitive} we illustrate 
the honeycomb lattice reciprocal space primitive cell and the $k$-point sampling scheme we have chosen.
In an effort to achieve a satisfactory compromise between
computational load and accuracy we, first of all, make use of the hexagonal symmetry inherent in the problem. 
This allows us to limit our calculations to $1 / 6$th of the total
Brillouin zone area when we distinguish $K$ and $K'$ valleys, and 
$1 / 12$th of the total area when we do not.
In addition, instead of using a uniform grid in the whole Brillouin zone we use a 
{$ k$}-point mesh which is denser near the Dirac point.
In our calculations we have used $18 \times 18$, $32 \times 32$ or $42 \times 42$ coarse grids for the full primitive cell 
and multiplication factors of up to $128$ for the finer $k$-point mesh near the valley centers.  The fine mesh 
densities were either $2304 \times 2304$ or $2048 \times 2048$ for typical calculations
at zero potential bias, and $672 \times 672$ when smaller densities were enough to converge the 
calculations in the case of strongly biased bilayers.
Dense $k$-point meshes were normally employed within $\sim 0.5/a$ of a Dirac point, and wider 
regions were used for strong bias cases.  (Here $a = 2.46 \AA$ is the lattice constant
of the triangular periodic lattice structure of graphene.)  Nevertheless $k$-point sampling 
approximations remain the main source of numerical inaccuracies.
The self-consistent-field calculations were iterated until convergence to 14 significant figures 
in the total energy per electron was achieved for a given choice of $k$-point sampling.
In order to resolve the small energy differences between solutions of the self-consistent field equations 
corresponding to different states 
we needed to compare results obtained with the same $k$-point sampling scheme.
The $k$-point sampling we used achieved convergence to within a few parts per thousand
for charge density differences and band gaps.

\section{Lattice Model Pseudospin Ferromagnet}
The inversion symmetry breaking instability in bilayer graphene 
occurs nearly independently at the two points and, within mean-field-theory,
entirely independently for each electron spin. \cite{pmag}
It is strongest in electrically neutral bilayers.\cite{hongki,fanzhang,yang,nandkishore1}
Initial studies carried out at the Hartree-Fock level \cite{pmag}
identified a family of competing pseudospin ferromagnet states.
Perturbative renormalization group calculations \cite{fanzhang,yang,fradkin} have 
confirmed that these instabilities can survive beyond mean-field-theory.
\begin{figure}[htbp]
\begin{center}
\includegraphics[width=6.6cm,angle=90]{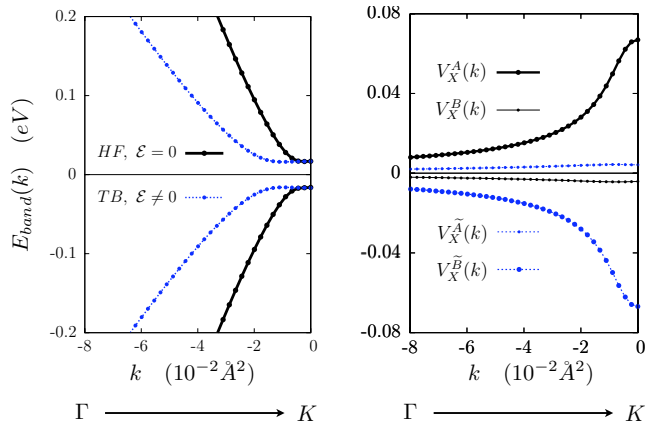} 
\caption{
{\em Left panel:}
Non-interacting biased tight-binding model (dotted line) and unbiased
Hartree-Fock (solid line)  
band structures for bilayer graphene for k moving away from the 
$\Gamma$ point towards the Dirac point $K$.
The broken symmetry state bands are compared with non-interacting 
electron bands with inversion symmetry explicitly broken by an externally applied
electric field ${\cal E} = 0.1 V/nm$.
The interacting system bands exhibit enhanced velocities and a
band gap due to 
spontaneously broken inversion symmetry.
These results were obtained for a model with hopping parameters 
$\gamma_0=-3.12 {\rm eV}$, $\gamma_1=-0.377 {\rm eV}$, 
$\gamma_3=\gamma_4 = 0$, and dielectric screening parameter $\epsilon_r = 4$.
{\em Right panel:}
Onsite $k$-dependent exchange potentials of the spontaneously broken symmetry state on the four bilayer sublattices.
The onsite potential is larger in magnitude on the 
low-energy sites that do not have an opposite layer neighbor, and 
lower on average in the layer with the larger density.  
}
\label{figure_pmag}
\end{center}
\end{figure}

Fig. \ref{figure_pmag} illustrates typical mean-field theory band structures for a 
$\pi$-band tight-binding model of unbiased bilayer graphene.
The non-interacting bands exhibit the $p^2$ dispersion at small $p$
which is captured by the massive chiral fermion model.  When interaction effects 
are included in the mean-field Hamiltonian, a gap opens up 
and velocities increase substantially.  The gap to the remote bands associated with the 
high energy states is also increased.  A gap at $p=0$ (where $\bf{k}=K$ so that $f({\bf k}$) vanishes) 
is always associated with a difference in mean-field atomic $\pi$-orbital energies between the two low-energy 
sites $A$ and $\tilde{B}$ and therefore a violation of the inversion symmetry 
which makes the two layers equivalent.  We refer to this broken symmetry state as a 
pseudospin ferromagnet, motivated by its continuum model description \cite{pmag}
in which the sublattice degree-of-freedom plays the role of a pseudospin.  
The increase in velocity at non-zero $p$ illustrated in Fig. \ref{figure_pmag} occurs even when inversion
symmetry is not broken\cite{polinissc}.
In the same figure we plot the
$k$-dependent on-site exchange potentials of the broken symmetry state 
which have been
obtained from the self consistent solutions.  The exchange potentials have
opposite signs on opposite layers,  
larger magnitude at the low energy sites $A$ and $\widetilde{B}$, 
and values that increase as the valley points are approached.

It is appropriate at this point to address some of the 
difficulties which arise in constructing reliably predictive theories for the influence of 
electron-electron interactions in graphene sheets.  In continuum model 
theories it is customary to introduce a relative dielectric constant 
$\epsilon_r \sim (\epsilon_{sub}+1)/2$ to account for dielectric screening due
to the substrate on which the graphene sheet lies.  ($\epsilon_{sub}$ is the 
dielectric constant of the substrate.  $\epsilon_{sub} \sim 4$ for 
substrates commonly used to support mechanically exfoliated graphene samples.  
This effect is important for some graphene sheet properties but is often
omitted in {\em ab initio} calculations.)
Some of the results we present below suggest that 
the continuum model is reliable, even for neutral 
graphene systems (which have less screening) and even when $\epsilon_r$ is small, as 
it should be in a model intended to describe suspended graphene samples and in models of graphene on a substrate with a 
small dielectric constant.  As we explain later, the two-band continuum model tends to be
less accurate for bilayer graphene than for single-layer graphene.
The implication for the lattice model
employed here, is that the on-site Coulomb interaction, determined by the charge form factor, 
will have a bearing on the model's predictions especially when $\epsilon_r$ is small.
Hartree-Fock mean-field theory, used in this paper to capture the non-local 
exchange properties which drive pseudospin ferromagnetism,
tends to overestimate the onset of broken symmetries.  In ab initio calculations 
it is common\cite{xalpha} to reduce the strength of bare exchange, say by factor of $\sim 2$,
to account for Coulomb correlation screening missing in an exchange only theory.  
This type of consideration may justify using a value of $\epsilon_r$ larger than 
the one which would be suggested by dielectric screening considerations alone, adding another 
level of uncertainty to any quantitative predictions.

\subsection{Total layer density and Chern number classifications of competing states}

\begin{table*}[htp] 
\caption{
Summary of Hall transport properties for valley $\tau_z = K,K'$ and spin $\sigma_z = \uparrow, \downarrow$ dependent layer polarization $\lambda_z=T,B$. 
These states are classified as layer  
ferromagnetic (F), layer ferrimagnetic (Fi) or layer antiferromagnetic (AF) following 
reference [\onlinecite{pmag},\onlinecite{xiaofan}].  F, Fi, and AF states have 
respectively four, three, and two valley-spin components polarized toward
the same layer ($\lambda_z = T$).
We have listed only eight out of a total of sixteen configurations omitting the equivalent configurations 
obtained by layer reversal for every flavor.  Each valley contributes a finite Hall conductivity with magnitude 
$\frac{e^2}{h}$ and a sign that reverses with both valley interchange and layer polarization.
The Hall conductivities are assigned to particular valleys on the basis of approximate Chern indices obtained by integrating 
Berry curvatures over the portion of the Brillouin zone near $K$ or $K'$.
The edge-state structure is expected to depend on each of the partial Hall conductivities. 
The total Hall conductivity is separated into contributions from separate spins and separate valleys using  
$\sigma_{xy}^{tot} =  \sigma_{xy}^{\uparrow} + \sigma_{xy}^{\downarrow}
= \sigma_{xy}^{K} + \sigma_{xy}^{K'}$.
The F configuration have zero total hall conductivity, whereas all Fi configurations 
have a finite total anomalous Hall conductivity of two units.
The $A$ configurations are expected to be electrostatically favored in the absence of an external 
layer bias potential and include three types of solutions with different 
Hall properties, one with a total anomalous total Hall conductivity,
one with zero Hall conductivity but finite spin Hall conductivity, 
and another with zero Hall and spin Hall conductivities because of opposite 
Hall conductivity contributions from $K$ and $K'$ valleys for both spins.
} 
{
\centering 
\begin{ruledtabular}
\begin{tabular}{l | c c c c | c c c c | c c | c c |  c |} 
\multicolumn{1}{c|}{ }        &
\multicolumn{4}{c|}{ $\left( \;\; \lambda_z \;\; \;\;\; \tau_z \;\; \;\;\;  \sigma_z  \;\; \right) $}        &
\multicolumn{1}{c} { $\sigma^{K, \uparrow}_{xy}$ }       &      
\multicolumn{1}{c} { $\sigma^{K', \uparrow}_{xy}$ }     &       
\multicolumn{1}{c} { $\sigma^{K, \downarrow}_{xy}$ }       &      
\multicolumn{1}{c|} { $\sigma^{K', \downarrow}_{xy}$ }     &  
\multicolumn{1}{c} { $\sigma^{\uparrow}_{xy}$ }       &      
\multicolumn{1}{c|} { $\sigma^{\downarrow}_{xy}$ }     &       
\multicolumn{1}{c} { $\sigma^{K}_{xy}$ }       &      
\multicolumn{1}{c|} { $\sigma^{K'}_{xy}$ }     &     
\multicolumn{1}{c} { $\sigma^{tot}_{xy}$ }   
         \\
\hline \hline                        
F &$ {(T \,\, K  \uparrow)}$  &  $ { (T \,\,  K'  \uparrow)}$  &  $(T \,\, K  \downarrow)$  &  $(T \,\, K'  \downarrow )$    &  1 & -1  &  1   & -1 &  0 & 0  &  2  & -2 & 0  \\
\hline \hline 
Fi &  $(T\,\,  K  \uparrow)$  &  $ (T \,\, K'  \uparrow)$  &  $(T \,\, K  \downarrow)$      &  $(B \,\, K'  \downarrow )$    &  1   & -1  &   1    &   1   &    0  &   2  & 2  &  0 &  2   \\
    &  $(T \,\, K  \uparrow)$  &  $ (T\,\, K'  \uparrow)$  &   $(T \,\, K'  \downarrow)$      &  $(B\,\,  K  \downarrow)$    &  1   & -1  &    -1  &  -1   &    0  &  -2  &  0 & -2 & -2   \\
    &  $(T \,\, K  \uparrow)$  &  $ (B \,\, K'  \uparrow)$   &  $(T\,\,  K  \downarrow)$      &  $(T \,\, K'  \downarrow)$    &  1   &  1  &    1    &   -1   &   2  &   0  &   2  & 0 & 2   \\
    &  $ (T \,\, K'  \uparrow)$  &  $(B \,\, K  \uparrow)$   &  $(T \,\, K  \downarrow)$      &  $(T \,\, K'  \downarrow)$    &  -1  & -1  &   1    &   -1   &   -2    &   0  &  -2  & 0  &  -2  \\
\hline \hline 
AF &$ (T \,\, K  \uparrow)$  &   $(B\,\,  K'  \uparrow)$    &  $(T \,\, K  \downarrow)$     &   $(B \,\, K'  \downarrow)$   &  1  & 1 &  1  &  1  &  2  & 2 &  2  &  2  & 4  \\
&$ (T \,\, K  \uparrow)$       &   $(B \,\, K'  \uparrow)$    &   $(T \,\, K'  \downarrow)$     & $(B\,\,  K  \downarrow)$     &  1  & 1  & -1  & -1  &  2  & -2 &  0  &  0  & 0  \\
&$ (T \,\, K  \uparrow)$       &   $(T\,\,  K'  \uparrow)$   &   $(B \,\, K  \downarrow)$     &  $(B\,\,  K'  \downarrow)$     &  1   & -1    & -1  &  1  &  0  & 0 &  0 & 0 & 0 \\
\hline 
\end{tabular} 
\end{ruledtabular}
}
\label{classify} 
\end{table*} 

Because each bilayer flavor 
can polarize toward either of the two layers, there are a total number of $2^4 = 16$ possible configurations
of the broken symmetry\cite{valleycoherence} state.
The layer polarization for each spin and valley determines the sign of the mass term
in its continuum model and has implications for the topological properties in the system as we will now discuss.
The 16 states can be classified\cite{pmag} by overall layer polarization as being 
ferromagnetic, ferrimagnetic, or layer antiferromagnetic.  By this classification\cite{pmag} there are two layer ferromagnetic states 
in which all flavors choose the same polarization, eight layer ferrimagnetic states in which
three of the four flavors choose the same polarization, and six layer antiferromagnetic states 
with no overall polarization.\cite{xiaofan}  The layer antiferromagnetic states are electrostatically favored in the absence of a potential bias. 

One of the most interesting properties of gapped bilayer graphene is the existence of a finite 
Hall conductivity and orbital magnetism due to the flavor-dependent momentum-space vortices\cite{nandkishore2,xiao,xiaofan} 
in the broken symmetry states.  Because the vorticity $v$ is opposite for opposite valleys,
the integrated Berry curvature gives rise to a Hall conductivity\cite{tknn,mingche,niureview,NagaosaReview} with magnitude 
$e^2/h$ for each flavor, 
and a sign that changes with valley as well as with layer polarization.
The Berry curvature reflects the handedness of Bloch electrons and captures 
intracell circulating currents which generate a finite
orbital magnetic moment 
proportional to the angular momentum 
due to self-rotating Bloch wave packets.\cite{mingche}
The Berry curvature of the system can be evaluated using\cite{niureview} 
\begin{eqnarray*}
\Omega_n \left( {\bf k} \right) = i \sum_{n^{\prime} \neq n} 
\left[ 
\frac{ \left< u_n \right|  \frac{\partial H}{  \partial k_x} \left| u_{n'}\right> 
\left< u_{n'} \right|  \frac{ \partial H}{ \partial k_y} \left| u_{n}\right>}
{ \left( E_{n^{\prime}} - E_{n} \right)^2} 
- c.c. \right]. 
\end{eqnarray*} 
where $\left| u_n \right>$ represents the Bloch eigenstates of the system and $E_n$ are the associated eigenvalues for each $k$.
The finite orbital moment generated by these wave packets has a similar expression and can be evaluated through\cite{niureview} 
\begin{eqnarray*}
m_n({\bf k}) &=& -(e/2m) L_n({\bf k})  \\
&=& -\frac{e}{2 \hbar}  i \sum_{n^{\prime} \neq n} 
\left[ 
\frac{ \left< u_n \right|  \frac{ \partial H}{ \partial k_x} \left| u_{n'}\right> 
\left< u_{n' }\right|  \frac{\partial H}{ \partial k_y} \left| u_{n}\right>}{E_{n^{\prime}} - E_{n} } 
- c.c. \right]. 
\end{eqnarray*}

\begin{figure}[htbp]
\begin{center}
\includegraphics[width=6.6cm,angle=90]{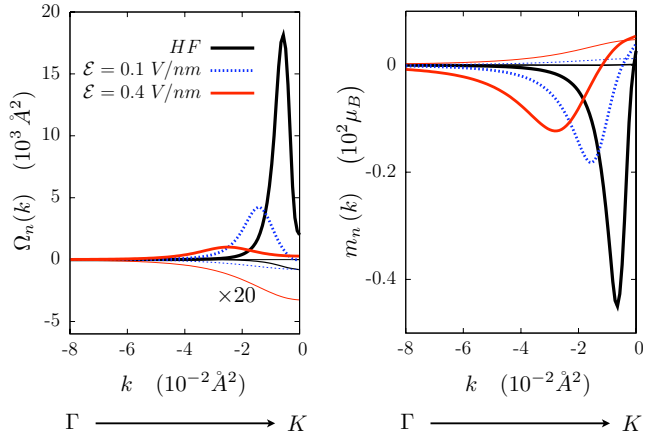} 
\caption{
{\em Left panel:}
Berry curvatures associated with the two valence bands for a Hartree-Fock (HF)
solution, and for non-interacting bilayer graphene in the presence of an interlayer bias.
We use a thicker line to represent the curvature of the low energy
band and a thinner line for the band farther from the Fermi level.
The small remote band contribution has been magnified $\times 20$. 
An interlayer electric field of ${\cal E} = 0.1 \, V/nm$ added to the non-interacting 
electron model gives a band gap comparable to the one
which emerges from our mean field calculations with $\varepsilon_r = 4$.
{\em Right panel:}
Orbital magnetization contributions from the two valence bands
in the vicinity of a valley point. 
We compare results for the self-consistent broken
symmetry states with those for non-interacting electrons with an
external interlayer bias.
}
\label{orbmag}
\end{center}
\end{figure}

A weak external magnetic field will tend to favor a state in which the orbital magnetizations of 
all valleys are aligned, and the anomalous Hall conductivity is correspondingly maximized.
This observation suggests
the possibility of valley optoelectronics that exploits the circular dichroism of 
interband transitions \cite{dichro} in the broken symmetry states.
The Kerr and Faraday effect measurements with linearly polarized light
can be a useful tool to detect signatures of broken time reversal symmetry.
\cite{kerr}
In Fig. \ref{orbmag} we present the $k$-dependent magnetization evaluated 
for one self-consistent gapped state in the presence of an interlayer bias.
Similar results have been obtained previously using the massive Dirac-fermion continuum model. 
An estimate of the zero field magnetization per valley-spin degree of freedom 
$M_{\tau_z \sigma_z} = \sum_n \int  m_n({\bf k}) \frac{d^2 {\bf k}}{\left( 2 \pi \right)^2}    
\simeq 2 \pi \sum_n \int   m_n(k) \, k \, \frac{dk}{\left( 2 \pi \right)^2}$ 
can be obtained integrating the orbital moment around each one of the valleys
for a given spin component. 
For the broken symmetry state with $\epsilon_r = 4$
each component integrates to $M_{\tau_z \sigma_z} =  10^{-3} \mu_B$ per carbon atom, compared to 
$M_{\tau_z \sigma_z} = 1.4 \cdot 10^{-3} \mu_{B}$ 
for the non-interacting system with bias ${\cal E} = 0.1 \,V/nm$ and 
$M_{\tau_z \sigma_z} = 2.2 \cdot 10^{-3} \mu_{B}$ per carbon atom for ${\cal E } = 0.4 \,V/nm$.
The individual flavor orbital magnetizations and  
anomalous Hall contributions cancel in states that do not have broken time-reversal symmetry.

In the non-interacting electron biased bilayer graphene state,
opposite contributions from the two valleys lead to vanishing total Hall conductivity,
consistent with the absence of broken time-reversal symmetry.
The nature of the layer polarized broken symmetry we describe here
is analogous to the biased bilayer in the sense that the charge transfer 
can be attributed to the ($k$-point dependent) 
exchange potential difference between low-energy sites on opposite layers, as illustrated in Fig. \ref{figure_pmag}.
Because of the non-locality of the exchange interactions, however, we are able to find 
self-consistent solutions in which the effective interlayer bias potential has 
opposite signs in the two valleys, and the total Hall conductivity is finite.  
In Table I we present a list of the 16 different configurations for which we have 
found self-consistent solutions, which are characterized by the 
sense of layer ($\lambda_z = T, B$) polarization for each valley ($\tau_z = K, K'$) and spin ($\sigma_z = \uparrow, \downarrow$)
as also discussed in reference [\onlinecite{xiaofan}].
These results suggest the interesting possibility of altering the quantum Hall 
conductivity of a graphene bilayer sample with an external bias potential as we discuss later.

\subsection{Broken inversion symmetry states in an unbiased bilayer}
We start by examining the layer antiferromagnetic charge balanced configurations which
are lowest in energy 
in the absence of an external bias 
because\cite{pmag} of the absence of a Hartree energy penalty. 
As summarized in Table I, there are three distinct types of layer antiferromagnets. \cite{xiaofan} 
One configuration is the anomalous Hall (AH) state, with four quantized units of Hall conductivity, 
in which electrons are polarized towards the same layer for both spins \cite{nandkishore2},
and towards opposite layers for opposite valleys.
Second is the spin Hall (SH) state which has opposite layer polarization on opposite
valleys, and in each valley opposite layer polarization for opposite spin.
The mean-field Hamiltonian for this state is similar to 
that of a non-interacting system with intrinsic spin orbit coupling \cite{prada}.  
Finally there is a solution with the same layer polarization in the two valleys, 
but opposite layer polarizations for opposite spins.  
This state has a finite valley Hall (VH) conductivity for each spin,
zero valley Hall conductivity when summed over spins, and 
zero total Hall conductivity \cite{xiao,morpurgo}.
Optical measurements based on linearly polarized light using the Kerr and Faraday 
effect can be used to identity the anomalous Hall state which has broken time 
reversal symmetry state.\cite{kerr}

\begin{table*}[htp] 
\caption{
Energy differences between distinct mean-field solutions in the absence of a potential bias.
The first group of columns represent energy differences
between the broken symmetry pseudospin ferromagnet states
in the anomalous Hall (AH) configuration and the unbroken symmetry state that has no gap.
The latter has been obtained constraining the self-consistent HF calculation to preserve
inversion symmetry. 
The third column represents the exchange energy difference $\Delta E_{X}^{TB}$ of the broken symmetry self-consistent state
with respect to the reference exchange energy of the non-interacting state.
The remaining columns represent energy differences between 
between anomalous Hall and valley Hall (VH) solutions.
The calculated energy differences depend sensitively on the choice of the model parameter 
$r_0$ used in the form factor.
The total exchange energy differences $\Delta E_{X}^{tot} = 4( \Delta E_{X}^{KK} + \Delta E_{X}^{K K'})$ 
can be expressed as a sum of intravalley ($KK$) and intervalley ($K K'$) 
contributions, where the four-fold factor is due to the twofold degeneracy in both spin and valley. 
We have verified that the energy differences depend very weakly on
the $\gamma_3$ and $\gamma_4$ parameters which are excluded in the minimal model.
The two anomalous Hall states have the same energy in mean-field theory as explained in the text.  
The values compiled in this Table have been evaluated using $\varepsilon_r = 4$ and the 
indicated carbon atom radii $r_0$. 
All energies are expressed in $eV$ per carbon atom unit.}
{
\centering 
\begin{ruledtabular}
\begin{tabular}{ |c|c|c|c|c|c|c|c|}   
\multicolumn{1}{c|}{}                          &
\multicolumn{3}{c|}{ $ E^{(AH/SH)} - E_{0} $}     &
\multicolumn{4}{c|}{ $ E^{(VH)} - E^{(AH/SH)} $ }     \\     \hline
$r_0 $   &   $\Delta E_{tot}$   &   $\Delta E_{X}$      & $\Delta E_{X}^{TB}$   &    $\Delta E_{tot}$  
    & $\Delta E_{X} $   &   $\Delta E_{X}^{KK}$   &   $\Delta E_{X}^{K K'}$   \\ \hline
$\widetilde{a}_0$  & $-4.84 \cdot 10^{-8}$ & $-1.13 \cdot 10^{-7}$ &     $-5.10 \cdot 10^{-4}$  &  
$ -1.03 \cdot 10^{-7}$    
& $-9.72 \cdot 10^{-7}$    & $-1.75 \cdot 10^{-7}$  & $-6.75 \cdot 10^{-8}  $    \\
2$\widetilde{a}_0$ & $-5.17 \cdot 10^{-8}$ & $-1.24 \cdot 10^{-7}$&   $-6.15  \cdot 10^{-4} $ &   
$-4.66 \cdot 10^{-9}$ 
&  $-1.92   \cdot 10^{-8}$ & $-4.09 \cdot 10^{-9}$ & $-6.97 \cdot 10^{-10}$ \\
$3\widetilde{a}_0$ & $-6.17 \cdot 10^{-8} $   & $-1.58 \cdot 10^{-7}$ &     $-5.09 \cdot 10^{-4}$   &  
$ -1.14 \cdot 10^{-9}$  & $-2.76 \cdot 10^{-9}$  & $-5.71 \cdot 10^{-10}$  & $-1.18 \cdot 10^{-10}$    \\
\end{tabular} 
\end{ruledtabular}
\label{edifs} 
}
\end{table*}

\begin{table*}[htp] 
\caption{
Transferred charge per valley-spin flavor in the charge 
balanced anomalous Hall and valley Hall solutions.
In our mean field Hartree-Fock calculations the VH
solutions have slightly larger band gaps and interlayer charge transfers than their anomalous Hall counterparts.
In this table, charge densities are in units of $10^{11} cm^{-2}$ and band gaps 
in $meV$ units.
We have used the tight-binding model parameters $\gamma_0 = - 3.12 eV$, $\gamma_1 = -0.377 eV$, 
$\gamma_3 = -0.29 eV$, and $\gamma_4 = -0.12 eV$.
The $\gamma_3$ parameter has only a marginal influence on the broken symmetry state.
The value chosen for $\gamma_4$ term tends to accumulate more electrons at the low 
energy sites in the bilayer but does not influence the strength of the broken symmetry. 
The results reported here were obtained using dielectric constant $\epsilon_r=4$ and 
carbon atom radius $r_0 = 3 \widetilde{a}_0$.  In the rightmost set of columns we show the results obtained 
when a smaller 
carbon atom radius $r_0 = 2 \widetilde{a}_0$ is used in the form factor calculation in Eq. (\ref{formfactor}). 
} 
\centering 
\begin{ruledtabular}
\begin{tabular}{|c| c c c c  | c  c c c  | c c c c c c  | c c c c |} 
\multicolumn{19}{c}{Anomalous Hall / Spin Hall   (AH/ SH)}   \\   \hline
\multicolumn{1}{c}  {}               & 
\multicolumn{4}{c|} {$\gamma_0$,   \; $\gamma_1$ }           &
\multicolumn{4}{c|}  {$\gamma_0$,    \; $\gamma_1$, \; $\gamma_3$}   &
\multicolumn{6}{c|}  {$\gamma_0$,    \; $\gamma_1$, \; $\gamma_3$, \; $\gamma_4$ }   &
\multicolumn{4}{c}  {$\gamma_0$,    \; $\gamma_1$,     $ \;\;\; r_0 = 2  \widetilde{a}_0$  }   \\
\hline \hline                        
    $\epsilon_r$       &   $\Delta n_l$        &   $\Delta n_s^A$    &    $\Delta n_s^B$   &   $\Delta_{gap}$ 
&  $\Delta n_l$       &   $\Delta n_s^A$   &   $\Delta n_s^B$    &    $\Delta_{gap}$   
&  $\Delta n_l$       &   $\Delta n_s^A$   &   $\Delta n_s^B$    &   $\Delta n_s^{\widetilde{A}}$    &  $\Delta n_s^{\widetilde{B}}$    
&  $\Delta_{gap}$  
& $\Delta n_l$   &   $\Delta n_s^A$   &   $\Delta n_s^B$    &    $\Delta_{gap}$    \\ [0.5ex]  
\hline 
4 & 0.52  &  0.81 &  -0.29 &  33  &  0.52 &  0.80 &  -0.28  &  31   &  0.52 & 5.67  & -5.15  & -4.59   & 4.07   & 31   & 0.47   & 0.72   & -0.25  & 31  \\ 
5 & 0.42  &  0.60 &  -0.18 &  23  &  0.42 &  0.60 &  -0.18  &  22   &  0.41 & 5.38  & -4.97  & -4.61   & 4.20   & 21   & 0.38   &  0.54  &  -0.16  &  22 \\ 
6 & 0.34  &  0.46 &  -0.12 &  17  &  0.34 &  0.46 &  -0.12  &  16   &  0.34 & 5.20  & -4.86  & -4.63   & 4.29   & 16   & 0.32   &  0.42  &  -0.10 &  16 \\ 
\hline  
\multicolumn{19}{c}{} \\
\multicolumn{19}{c}{Valley Hall  (VH)}
\\  \hline
\multicolumn{1}{c}  {}               & 
\multicolumn{4}{c|} {$\gamma_0$,   \; $\gamma_1$ }           &
\multicolumn{4}{c|}  {$\gamma_0$,    \; $\gamma_1$, \; $\gamma_3$}   &
\multicolumn{6}{c|}  {$\gamma_0$,    \; $\gamma_1$, \; $\gamma_3$, \; $\gamma_4$ }   &
\multicolumn{4}{c}  {$\gamma_0$,    \; $\gamma_1$,     $ \;\;\; r_0 = 2  \widetilde{a}_0$  }   \\
\hline \hline                        
    $\epsilon_r$       &   $\Delta n_l$        &   $\Delta n_s^A$    &    $\Delta n_s^B$   &   $\Delta_{gap}$ 
&  $\Delta n_l$       &   $\Delta n_s^A$   &   $\Delta n_s^B$    &    $\Delta_{gap}$   
&  $\Delta n_l$       &   $\Delta n_s^A$   &   $\Delta n_s^B$    &   $\Delta n_s^{\widetilde{A}}$    &  $\Delta n_s^{\widetilde{B}}$    
&  $\Delta_{gap}$  
& $\Delta n_l$   &   $\Delta n_s^A$   &   $\Delta n_s^B$    &    $\Delta_{gap}$    \\ [0.5ex]  
\hline 
4 & 0.52  &  0.84 &  -0.32  &  33  &  0.52 &  0.80 &  -0.28  & 31   &  0.52 &  5.70   & -5.18  & -4.55   & 4.03   & 31   & 0.49   & 0.89   & -0.40   &  32    \\ 
5 & 0.42  &  0.61 &  -0.19  &  23  &  0.42 &  0.61 &  -0.19  & 22   &  0.42 &  5.40   & -4.98  & -4.60   & 4.18   & 22   & 0.40   & 0.64  &  -0.24   &  22     \\ 
6 & 0.34  &  0.47 &  -0.13  &  17  &  0.34 &  0.47 &  -0.13  & 16   &  0.34 &  5.21  &  -4.87  & -4.62   & 4.28   & 16   & 0.33   & 0.48   &  -0.15  &  17     \\ 
[1ex] 
\hline 
\end{tabular} 
\end{ruledtabular}
\label{hoppings} 
\end{table*} 

In the continuum model formulation the three distinct solutions are degenerate. 
In the lattice model Hartree-Fock theory the energies of the AH and SH states
are still exactly degenerate because the exchange energy is spin
diagonal and the energy within each spin is independent of the Hall effect sign.
We refer to these two states collectively as the anomalous Hall states.
The valley Hall (VH) solution does have a different energy however, because 
the relative sense of layer polarization in the two valleys influences inter-valley exchange potentials.
Table \ref{edifs} presents our results for the condensation energy of the broken symmetry state and 
for the differences in energy between the VH and AH/SH solutions separated into different contributions.
We first note that the condensation energies are reasonably independent of the model 
parameters $\epsilon_r$ and $r_0$, which are not precisely known and dependent on the 
sample's dielectric environment.  The scale of the condensation energy 
should be compared with the value of the Coulomb interaction at the momentum scale $k^*$ ($e^2 k^*$) 
over which the two-band model applies to bilayer graphene.  We find $k^*$ by 
setting the low-energy model parabolic band energy equal to the isolated layer energy, 
which gives $k^* \sim \gamma_1/\hbar v$.  The Coulomb energy at this momentum scale is 
$\sim (e^2/\epsilon_r \hbar v) \times \gamma_1 = \alpha_{gr} \times \gamma_1 \sim 100 {\rm meV}$.
This energy scale gives an estimate of the size of the gap which would be expected if 
all states with $k < k^*$ were fully layer polarized.  We find a gap which is approximately
four times smaller and density shifts between layers that are smaller than 
$\pi k^{*2} /(2\pi)^2 \sim (\gamma_1/2 \pi W a)^2 $ by a similar fraction.
These numerical values indicate that the broken symmetry arises mainly from those 
bilayer band states that are reasonably well described by the low-energy
two-band models used in the original mean-field calculations, as expected. 
We note that the condensation energy is smaller by several orders of magnitude than the 
total interaction energy, which involves electrons across the full $\pi$-band. 

The second block of columns present differences in energy between VH states and AH/SH states.
For the parameters we have considered the total energies 
for the VH states are lower than the AH/SH thanks to the exchange energy advantage of the former
states.
The differences in energy between these 
states become substantially smaller than the condensation energy
when the carbon radius model parameter $r_0$ is increased.  Larger values of $r_0$ improve the 
accuracy of the continuum model and differences in energy between the different solutions 
rapidly decrease.  We have previously argued that values of $r_0 \sim 3 \widetilde{a}_0$
in the form factor are appropriate.  If so, the differences in energy 
between different solutions are $\sim  10^{-9} {\rm eV}$ per carbon atom,
about 100 times smaller than the ordering condensation energy.
We have also explored an alternative formulation of the 
Hartree-Fock equations which evaluates the non-local exchange potentials in real space. 
This formulation allows the model's onsite repulsion $U$ to be adjusted separately 
from the longer ranged tails\cite{bfieldinsulator,latticescale} and thus
allow a more direct assessment of the impact of lattice scale details of the effective Coulomb interaction,
and leads to similar conclusions.  

Large values of $U$ imply strong short-range correlations which are not accurately described by the 
continuum model.
From the results shown in table \ref{edifs}, however,
we see that only moderate short-range screening from degrees-of-freedom 
outside the $\pi$-band model are required to make deviations from the 
continuum model small.  Given the abundance of separate evidence that 
interaction effects in graphene systems are accurately described by a 
continuum model, we assume that the required short-range screening is in fact present.

In Table \ref{hoppings} we report on the layer 
distribution of charge when inversion symmetry breaking occurs within a particular valley.
The results presented here are single valley-spin results based on the AH and VH solutions.
We have performed these calculations for band models which include and exclude tight-binding hopping 
parameters other than in plane
nearest neighbor hopping $\gamma_0$ and interlayer tunneling $\gamma_1$.
The trigonal warping $\gamma_3$ term connects the sites $A$ and $\widetilde{B}$ 
and introduces a triangular distortion in the band structure 
near the Dirac point, hence breaking the approximately circular symmetry of the bands. 
It also makes the dispersion become linear instead of quadratic at the lowest energy scales 
and works against symmetry breaking. 
We notice that the reduction of the band gap due to trigonal warping varies between two per cent 
and ten per cent depending on the short-range interaction strength, indicating that trigonal warping plays a relatively minor
role.  The $\gamma_4$ term increases the accumulation of charge density at lattice
sites $A$ and $\widetilde{B}$ relative to that at the high energy sites $B$ and $\widetilde{B}$.
The gap size and total transferred charge ($\Delta n_l$) 
remain virtually unchanged, although the distribution of charge between sublattices on the same layer ($\Delta n_s^i$)
is altered in some states.  
\begin{figure}[htbp]
\begin{center}
\includegraphics[width=8cm]{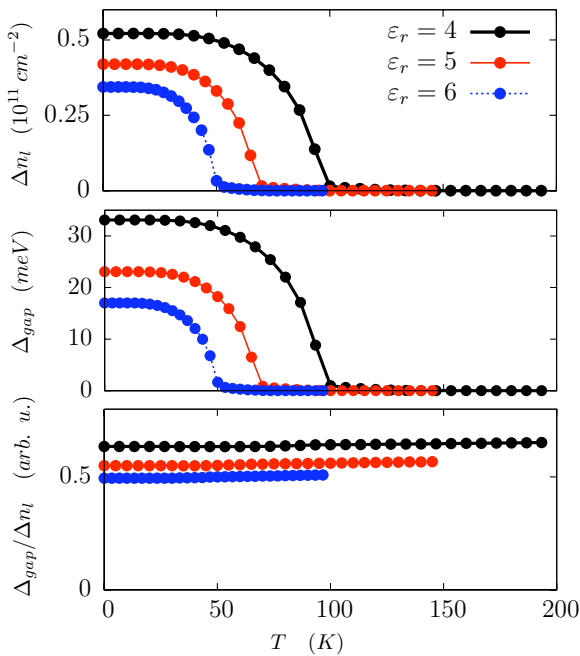} 
\caption{
Temperature dependence of mean-field-theory charge transfer per valley-spin and the associated band gap. 
{\em Top panel:}
Total charge per unit area ($10^{11} \,\, cm^{-2}$) transferred from one layer to another $\Delta n_l$  per each valley-spin
polarized component  
as a function of temperature. We represent the dependence for different values of the relative dielectric constant $\epsilon_r$.
{\em Middle panel:}
The bottom panel presents the mean-field-theory band gap as a function of temperature. 
{\em Bottom panel:} 
Ratio between the band gap and transferred charge density.
}
\label{tempdep}
\end{center}
\end{figure}

The temperature dependence of the band gap and the transferred charge per valley-spin component
in mean-field-theory is plotted 
in Fig. \ref{tempdep}.  As the temperature is increased both the charge density and band gaps are reduced,
but their ratio is approximately fixed. 
The decay trend is similar for the different dielectric constants we have considered.
The critical temperatures obtained in this 
mean-field calculation provide an 
estimate of the maximum temperature at which local order survives in the 
absence of disorder.    
The band gap decreases more quickly than the charge-density transfer order 
parameter as the interaction is made weaker;  
the charge density per valley-spin varies approximately like 
$\varepsilon_r^{-1}$ whereas the gap varies as $\varepsilon_r^{-1.55}$.

\subsection{Quantum phase transitions in the low magnetic field and low bias regime}
To first order in magnetic field, the magnetic contribution to energy is simply proportional to 
spontaneous magnetization discussed previously, which is entirely orbital in character and 
depends on the flavor-dependent layer polarizations.  
States that are related by layer polarization reversal are no 
longer equal in energy in the presence of a magnetic field.   
If we take $\Delta E_{tot} \sim 10^{-9} \, eV$ per carbon atom for 
the mean field energy differences between AH and VH states from Table \ref{edifs}
and use the relation $\Delta E_{tot} \sim   M_{\tau_z \sigma_z} \cdot B_c$,
we obtain from the orbital magnetization values we have calculated 
that a magnetic field $B_c \sim 0.004 T$ is sufficient to favor the AH state with 
orbital magnetization parallel to the magnetic field
over VH states.  Considering the small energy differences between
the different competing states as shown in table \ref{edifs},
we can expect that the coupling between magnetic field and the orbital moment in the system
can play a decisive role in selecting the minimum energy ground-state.
Because the electron densities at which gaps occur are magnetic field 
dependent in states with finite Hall conductivity, 
small fluctuations in the density like those associated with electron-hole puddle domains \cite{enrico}
will also have an important influence on the nature of the ground-state at low magnetic fields.

\begin{figure}[htbp]
\begin{center}
\includegraphics[width=8.3cm]{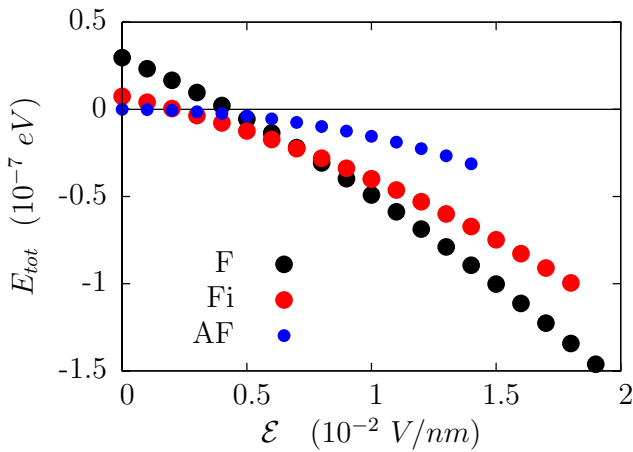} 
\caption{
Bias dependence of the total energies of the system for different 
valley-spin polarization configurations for a system with $\varepsilon_r = 4$. 
The system undergoes transitions from the AF to Fi and then to F states as a function of external field, 
each one displaying clearly different quantum Hall conductivity properties. 
The origin of energy has been arbitrarily shifted for presentation convenience.
}
\label{lowbias}
\end{center}
\end{figure}

\begin{figure}[htbp]
\begin{center}
\includegraphics[width=8.5cm,angle=0]{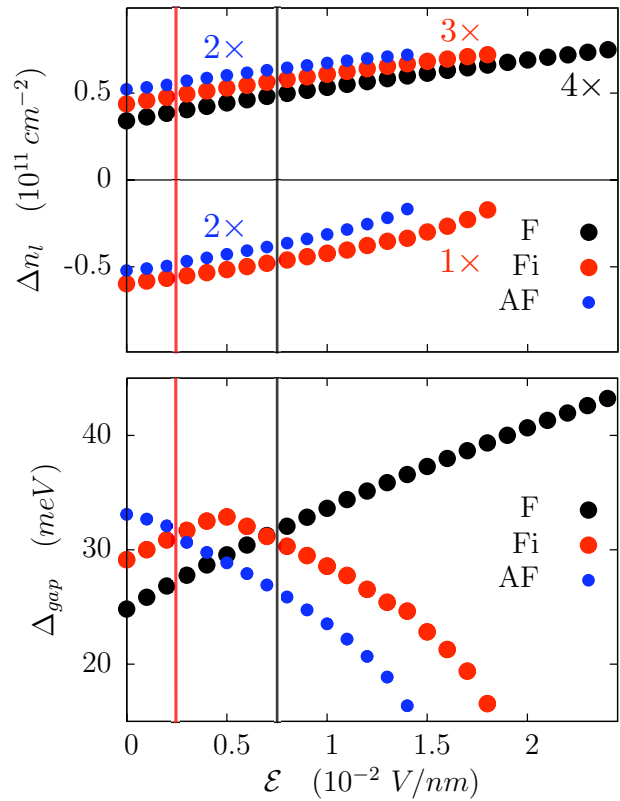} 
\caption{
{\em Upper panel:}
Each branch represents the bias dependence of the valley resolved layer polarization density 
associated with one valley-spin component.  These results were obtained for a system with $\varepsilon_r = 4$. 
The multiplication factors represent the number of times each branch needs to be repeated
to give the total charge density to complete the contributions from the four components.
{\em Lower panel:}
Band gap as a function of bias for F, Fi and AF states for a system with $\varepsilon_r = 4$.
We observe a competition in the band-gap opening between exchange and 
Hartree contributions which can lead to a
reduction of the band gap as the bias is increased.
}
\label{lowgap}
\end{center}
\end{figure}

We now consider the case of zero magnetic field and finite electric field.
In presence of an external bias the inversion symmetry of a graphene bilayer is explicitly broken,
favoring charge accumulation in one of the layers.
When the external bias becomes large enough the charge balanced broken symmetry antiferromagnetic (AF) 
configuration gives way to ferrimagnetic (Fi) and eventually ferromagnetic (F) configurations.
These solutions can become energetically favored for certain ranges of the external bias potential
thanks to their intrinsic broken inversion symmetry configuration with spontaneous charge transfer.
In Fig. \ref{lowbias} we present the total energies of the different types of solutions in the low bias regime,
obtained by starting the self-consistent calculations from different seeds.
For low potential bias the charge balanced AF structure remains lowest in energy.
Eventually the external field becomes large enough to flip the layer polarization of one valley-spin
component towards the layer favored by the electric field, giving rise to the Fi type solutions.
The electrostatic charge imbalance of the Fi states is approximately half of that associated with the F state
in which all
four components are polarized towards the same layer, 
a fact that can be inferred from the slopes of the total energy evolution as a function 
of external electric field.
In Fig. \ref{lowgap} we present the charge densities associated with one valley-spin component
with different layer polarizations.
For the AF and Fi solutions we have components that are polarized toward both layers
that we represent as $\Delta n_l = (n_l - n_0)$, the density difference with respect to the uniform background density $n_0$.  
In the Fi configuration the three charge density components polarized towards the same layer do not have exactly the 
same value when the electron spins are different, but they are similar in magnitude.
In the F configuration all four density components are polarized towards the same layer and have the same magnitude.

The behavior of the band gaps in the low bias regime illustrates the character of each solution.
The AF band gap continually decreases when the electric field is increased. 
In this case the electric field always works against the charge density distribution 
of two components directly contributing in the reduction of the band gap.
For the F configuration we have a completely opposite trend; the gap always increases
since the electric field favors charge
accumulation in the layer in which the density is already higher. 
In the Fi configuration we find an intermediate situation.
We find an initial increase of the gap in presence of an electric field thanks to the 
reduction of the Hartree energy penalty associated to the spontaneous charge imbalance. 
Beyond a certain point the band gap starts to decrease, reflecting the fact that 
the electric field is working against the layer polarization of one component. 
Eventually when the effect of the external field becomes dominant  the system undergoes a phase
transition to the F configuration. 
Every time there is a crossover of the minimum energy levels
we will see an abrupt change in the band gap and charge density.
The Hall conductivity of the system will also see discontinuous changes 
following the classification of table \ref{classify}.
The observations above suggest that the band gap will show a non-monotonic dependence on potential bias 
with an initial reduction at low fields for AF states and an increase for large enough bias for F states.
Between those two limits the system can form a Fi state that also 
has a band gap that decreases with increasing electric field,
before finally making a transition to the F state.

\subsection{Exchange screening effects in presence of a strong bias}
\begin{figure}[htbp]
\begin{center}
\includegraphics[width=8cm]{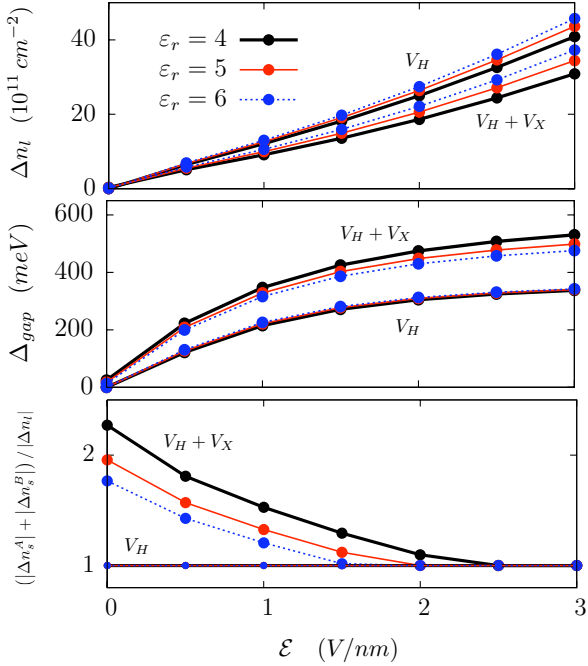} 
\caption{
External bias dependence of layer resolved charge density, band gaps and sublattice charge asymmetry.
We have considered the F configuration where all four valley-spin components are polarized towards the same layer. 
{\em Upper panel:}
Charge accumulation dependence as a function of bias in presence of Hartree and Fock terms
and charge accumulation dependence as a function of bias in presence of Hartree screening only.
The evolution of $\Delta n_l$ as a function of bias has a slightly larger slope in the Hartree only approximation 
than the Hartree-Fock case.
{\em Middle panel:}
Band gaps obtained for Hartree-Fock and Hartree approximations
using different values of $\varepsilon_r$ as a function of external electric field ${\cal E}$.
We notice a substantial enhancement of the band gap when the exchange term is included.
{\em Lower panel:}
Measure of deviation from neutrality in the charge differences for each one of the 
sublattices in a given layer normalized by the total sloshed charge.
The asymmetric distribution of the charge between low energy sublattice $A$ 
and high energy sublattice $B$ of a given graphene layer is enhanced in presence 
of electron exchange. This asymmetry becomes smaller as the external bias
becomes stronger.
}
\label{strongbias}
\end{center}
\end{figure}

In the presence of an external bias the inversion symmetry of graphene bilayer is explicitly broken,
favoring the charge accumulation in one of the layers and opening a band gap, a fact that has been
verified in several experiments \cite{ohta,delft, yuanbo} and also predicted theoretically within tight-binding
\cite{castro},
tight-binding plus Hartree screening \cite{mccann,zhangscr} or ab-initio calculations \cite{hongki}.
Here we explore the role that electron exchange can play when 
the system is subject to a strong external bias.
In Fig. \ref{strongbias} we present the charge accumulation in one of the layers as a function 
of bias. In the strong bias limit we are in the F 
configuration in which all four components are polarized towards the same layer. 
Results obtained in the Hartree approximation follows a similar trend in the strong bias limit resulting
in comparable amounts of total transferred charge in the range of bias we considered, 
with the Hartree only screening allowing more sloshed charge. 
A larger value of dielectric screening $\varepsilon_r$ that weakens the interaction strength
also weakens the electrostatic screening and therefore more charge imbalance per layer is expected.

As we show in Fig. \ref{strongbias} the presence of the exchange term in the
Hamiltonian introduces a clear enhancement of the band gap that persists up to high bias potentials.
This gap enhancement can be related to the exchange contribution that tends to 
introduce a strong asymmetry in charge distribution between $A$ and $B$ sublattices
of a given layer that persists up to rather high values of external electric field as shown
in Fig. \ref{strongbias}.
We may define this intralayer charge imbalance ratio as
$\left( \left| \Delta n_s^A \right| + \left| \Delta n_s^B \right| \right) / \left| \Delta n_l \right|$.
The enhancement of charge imbalance between the sublattices is one of the effects of 
intralayer exchange that contributes to changing the magnitude of the band gap.
This quantity is more sensitive to the strength of the Coulomb interaction 
than to changes in the temperature.

\section{Summary and conclusions}
We have presented a numerical description of the broken charge symmetry 
Hartree-Fock ground state in a graphene bilayer system for different dielectric
screening of the media and temperature. 
The predicted gap sizes and temperature ranges in which this effect is expected to survive
should be experimentally accessible.  The size of the interlayer 
charge transfers suggests that the broken symmetry states will be suppressed
by smooth disorder strong enough to produce charge puddles with density variations
larger than around $10^{-5}$ per carbon atom.
The total amount of spontaneous charge transfer from one layer to another in the broken 
ground state is $\sim  10^{11} cm^{-2}$, 
smaller but comparable in magnitude to the densities that can be induced by 
gating the device or depositing impurities 
\cite{geim} and can therefore be relevant for interpreting gating experiments in bilayer graphene. 
The mechanism for broken symmetry described here is 
most effective for a neutral bilayer because 
the energy gain due to charge transfer is greatest for electrons nearest to the
Dirac point. 
The broken symmetries are expected to be present only when the carrier density is smaller than about $10^{11} cm^{-2}$,
hence a reduction of this  
effect is expected as the Fermi surface moves away
from the charge neutrality point when the system is doped.

The Coulomb correlation neglected in the
Hartree-Fock formalism usually plays a more important in extended systems
\cite{2dmontecarlo} than in localized atomic systems
where they are typically one order of magnitude smaller than exchange effects.
Treatments of Coulomb correlation based on a renormalization group framework \cite{fanzhang} 
or RPA \cite{nandkishore1} suggest that the broken symmetry addressed in this paper 
will survive when Coulomb correlation is considered. 
In our case we have used larger values of $\varepsilon_r$ 
than those based on estimates from dielectric functions alone 
in order to effectively reduce the strength of exchange  
to approximately account for screening effects.
Application of a self-consistent dynamic screening approximation to this 
problem would be interesting but would require the 
high density of sampling $k$-points near the Dirac point to be 
implemented in a computationally efficient way.

An interesting feature of these broken symmetry states is the spontaneous quantum 
Hall effects which appear when opposite valleys have opposite layer polarizations.\cite{xiaofan}
When opposite valleys have the same layer polarization, the system has only a valley Hall effect, which is not manifested in standard electrical measurements.  There is no energy difference between these states in a continuum model which does not account for lattice-scale physics.  We find that the energy difference between these states decreases rapidly depending on the strength of the on-site effective interaction in our $\pi$-band only model, with valley Hall states being favored over anomalous Hall states.  For a very weak on site interaction, anomalous Hall states could be favored over valley Hall states.  States in which electrons with one spin orientation form one Hall state, while electrons with the other spin form the other Hall state also occur.  
Despite the uncertainties of the calculation due to this high sensitivity to lattice details
they do give us an estimate for the magnitude of the energy differences between different configurations.
Unlike Nandkishore and Levitov, \cite{nandkishore2} we find that valley Hall states have lower energy than anomalous Hall states, but that the energy competition can be altered by even quite weak external magnetic fields.  This scenario appears to be consistent with experiments from the Yacoby group in which a $\nu=4$ quantized Hall effect persists down to very weak magnetic fields.   The small energy differences between competing states could mean that domains with all characters are present, separated by domain walls,  because of entropic and disorder considerations.  In the presence of a magnetic field, coupling of spontaneous orbital magnetism to the external field favors anomalous Hall states and should coarsen any domain structure.  Similarly an externally applied potential bias favors valley Hall states and should also coarsen domain structures.   In the absence of disorder we do find that the charge gap is first decreased by potential bias, and finally increased once both spins have valley Hall effects with the same layer polarization.  This finding is qualitatively consistent with experiment.   
\cite{yacoby_latest}

We have also presented mean field theory estimates of the critical temperatures associated with spontaneous layer polarization states in graphene bilayers.
According to these estimates,  order will not survive to room temperatures and therefore is unlikely to be useful for applications.  On the other hand, the exchange interactions which produce broken symmetry states at low temperatures, will still play a role in enhancing gaps produced by external 
potential biases at room temperature.

\subsection*{Acknowledgments}
We gratefully acknowledge discussions with Hongki Min, Marco Polini, Byounghak Lee, Wang Yao,
Rahul Nandkishore, Dagim Tilahun and Andrea Young.
Financial support was received from Welch Foundation grant TBF1473, 
NRI-SWAN, DOE grant Division of Materials Sciences and Engineering DE-FG03-02ER45958.


\begin{thebibliography}{99}
\bibitem{valleycoherence} For unbiased bilayers, states in which 
layer polarization occurs in combination with coherence between different 
valleys are degenerate with states in which various layer polarizations 
occur within a valley.  In a lattice model, coherence between valleys implies 
a charge-density-wave structure in the broken symmetry state which is 
not permitted in the class of states we examine.  Although we believe that these 
states will have higher energies because of associated electrostatic energy 
increases, particularly so when an interlayer bias is present, they should be 
examined more closely in future work.
\end{thebibliography}
\end{document}